%% file: aaai24.tex
\documentclass[letterpaper]{article} 
\usepackage{aaai24}  
\usepackage{times}  
\usepackage{helvet}  
\usepackage{courier}  
\usepackage[hyphens]{url}  
\usepackage{graphicx} 
\usepackage{natbib}  
\usepackage{caption} 
\usepackage{algorithm}
\usepackage{algorithmic}
\usepackage{amsfonts}

\usepackage{newfloat}
\usepackage{listings}
\DeclareCaptionStyle{ruled}{labelfont=normalfont,labelsep=colon,strut=off} 
\floatstyle{ruled}
\newfloat{listing}{tb}{lst}{}
\floatname{listing}{Listing}
\usepackage{xcolor}
\usepackage{booktabs}

\title{WEIRD ICWSM: How  Western, Educated, Industrialized, Rich, and Democratic is Social Computing Research?}

\author{
    Ali Akbar Septiandri, Marios Constantinides, Daniele Quercia
}
\affiliations{

    Nokia Bell Labs \\
    Cambridge, UK \\
    \{ali.septiandri, marios.constantinides, daniele.quercia\}@nokia-bell-labs.com

}

\usepackage{bibentry}

\begin{document}

\maketitle

\begin{abstract}
Much of the research in social computing analyzes data from social media platforms, which may inherently carry biases. An overlooked source of such bias is the over-representation of WEIRD (Western, Educated, Industrialized, Rich, and Democratic) populations, which might not accurately mirror the global demographic diversity. We evaluated the dependence on WEIRD populations in research presented at the AAAI ICWSM conference; the only venue whose proceedings are fully dedicated to social computing research. We did so by analyzing 494 papers published from 2018 to 2022, which included full research papers, dataset papers and posters. After filtering out papers that analyze synthetic datasets or those lacking clear country of origin, we were left with 420 papers from which 188 participants in a crowdsourcing study with full manual validation extracted data for the WEIRD scores computation. This data was then used to adapt existing WEIRD metrics to be applicable for social media data. We found that 37\% of these papers focused solely on data from Western countries. This percentage is significantly less than the percentages observed in research from CHI (76\%) and FAccT (84\%) conferences, suggesting a greater diversity of dataset origins within ICWSM. However, the studies at ICWSM still predominantly examine populations from countries that are more Educated, Industrialized, and Rich in comparison to those in FAccT, with a special note on the 'Democratic' variable reflecting political freedoms and rights. This points out the utility of social media data in shedding light on findings from countries with restricted political freedoms. Based on these insights, we recommend extensions of current ``paper checklists'' to include considerations about the WEIRD bias and call for the community to broaden research inclusivity by encouraging the use of diverse datasets from underrepresented regions.
\end{abstract}

\input{sections/1_Introduction}
\input{sections/2_RelatedWork}

\input{sections/3_Methodology}

\input{sections/4_Results}

\input{sections/5_Discussion}

\input{sections/6_Conclusion}

\bibliography{aaai24}

\end{document}

%% file: sections/1_Introduction.tex
\section{Introduction}
\label{sec:introduction}
Social computing research, prominently featured at conferences such as the ICWSM, often analyzes data from social media platforms (e.g., Reddit, X) to understand human dynamics and inform the design and use of information in communication technologies that consider the social context. This area of study confronts challenges related to data inclusivity, primarily due to biases inherent in the data collection process, which often overlooks minority groups as a result of historical discrimination, selection practices, and sampling methods~\cite{olteanu2019social, baeza2018bias}. Another bias is introduced by the predominant focus on WEIRD populations, an acronym denoting Western, Educated, Industrialized, Rich, and Democratic societies~\cite{henrich2010weirdest}. It describes research that is centered around these groups but lack a comprehensive global perspective, potentially skewing the depiction of human behavior and leading to conclusions that might not accurately represent the broader human experience~\cite{Heckman2010}.

Recent research highlights an inclination towards Western and particularly US-centric viewpoints in academic studies. Psychology studies mostly analyze data from WEIRD populations, comprising just 12\% of the global population~\cite{henrich2010weirdest}. Similar findings were also reported in computing conferences, publishing human-computer interaction (HCI) and responsible AI research. A meta-analysis of the ACM CHI conference papers from 2016 to 2020 showed an overwhelming 73\% of studies focusing on Western populations~\cite{linxen2021weird}, while an analysis of the ACM FAccT conference papers between 2018 and 2022 reported a staggering 84\% of studies focusing on Western populations, with 63\% of them drawing exclusively from the US~\cite{septiandri2023weird}. Additionally, another analysis of the ACM FAccT proceedings spanning 2018 to 2021 raised alarm over the prevalent biases within word embeddings and computer vision technologies~\cite{laufer2022four}. Taken these together, it is important therefore to understand the degree to which published papers across scientific communities depend on WEIRD samples, and highlight the need for diversifying research perspectives.

In this study, we examined the extent to which ICWSM papers draw from WEIRD populations, and compared our findings with those reported in previous literature from two major computing conferences; one that focuses on human-centered design (ACM CHI) and another that focuses on democratization and responsibility of AI (ACM FAccT). We chose to focus on ICWSM instead of collating social media papers from variety of conferences because of two main reasons related to relevance and allowing for standardized comparison: 1)  ICWSM attracts a dedicated community of researchers and practitioners in web and social media research; and 2) Unlike broader conferences that cover a wide range of topics, ICWSM's specialization ensures that the research papers presented  are comparable with each other. In so doing, we made three main contributions:

\begin{enumerate}
    \item We collected 494 ICWSM papers published between 2018 and 2022, including full research papers, dataset papers, and posters. We conducted a crowdsourcing study with full manual validation in which 188 crowdworkers extracted information (e.g., number of participants, social media platform used) from these papers that allowed us to compute the WEIRD scores.\footnote{We made our annotated dataset publicly available for reproducibility \url{https://anonymous.4open.science/r/weird-icwsm-3638}}
    Upon removing papers that analyzed purely synthetic datasets or those where the countries of origin could not be identified, we were able to annotate and keep 420 papers for our analysis.
    \item Upon the annotated papers, we computed a set of five WEIRD metrics adopted from previous literature, and extended them to be applicable for social media data.
    \item Our analysis revealed that 37\% of ICWSM papers focus on datasets that exclusively represent Western countries. This proportion is significantly lower than the percentages observed at CHI (76\%) and FAccT (84\%), which may reflect the broader diversity of datasets used, including those from both Western and non-Western regions. Yet, the research presented at ICWSM generally examines populations from countries that are more Educated, Industrialized, and Rich compared to those in FAccT studies. A closer look at dataset and poster papers showed that they scored lower on the ``EIRD'' (Educated, Industrialized, Rich, and Democratic) metric than full research papers, suggesting that dataset and poster papers draw from populations of less Educated and Democratic countries. Additionally, we observed that papers with cross-country authorship tend to focus on samples from countries with lower levels of democracy. This suggests that studying social media data helps uncover findings from less democratic countries.
\end{enumerate}

We conclude with practical strategies to reduce reliance on WEIRD populations, including initiatives for broadening the scope of paper checklists, the inclusion of responsible AI statements, and promotion of authors' diversity. 

%% file: sections/2_RelatedWork.tex
\section{Related Work}
\label{sec:related}
WEIRD, an acronym for \textbf{W}estern, \textbf{E}ducated, \textbf{I}ndustrialized, \textbf{R}ich, and \textbf{D}emocratic, denotes a specific segment of the global population that is disproportionately represented in research studies, particularly within psychology and other social sciences~\cite{henrich2010weirdest}. WEIRD populations often come from countries such as the US, Canada, Western Europe, and Australia, which are characterized by higher standards of education, industrialization, affluence, and democratic systems than the global norm. \citet{henrich2010weirdest} study showed that 96\% of research participants in psychology are drawn from WEIRD populations and account only for the 12\% of the global population. 

Similar findings have been echoed in the computing field, with previous studies analyzing the proceedings of premier computing conferences that advocate for mitigation strategies for sampling issues. An analysis of 3,269 papers published at the ACM CHI conference between 2016 and 2020 showed that 73\% of them focused on Western populations~\cite{linxen2021weird}, mainly from US, Ireland, and Switzerland. A similar analysis of FAccT papers published between 2018 and 2022 showed that 84\% of these papers focused on Western populations, with two-thirds particularly from the US~\cite{septiandri2023weird}. Meta-studies also compared and contrasted these two conferences. An analysis of 200 papers from CHI and FAccT revealed a broad spectrum of participant roles, predominantly focused on authors and participants from the US~\cite{van2023methodology}. Additionally, \citeauthor{laufer2022four} undertook a reflexive meta-study of four years of FAccT proceedings, identifying key research themes such as group-level fairness and disinformation, while also exploring the community's priorities, including the demand for transparency in the peer review process and concerns about the influence of industry on research publications.

Despite the over-reliance on WEIRD populations, the computing field has increasingly begun to address such biases. Studies now explore cultural differences in product design~\cite{niess2021attitudes, busse2020cash, wilkinson2022many, ma2022enthusiasts}, the inclusion of minority views in computing education~\cite{oleson2022decade}, and digital accessibility in the Global South~\cite{nourian2022digital}. For example, research on fitness trackers showed that Arab users view them as advisory tools, contrasting with users from WEIRD populations who see them as prescriptive~\cite{niess2021attitudes}. Another study showed cultural variations in smartphone privacy expectations, highlighting that in places such as India, Pakistan, and Bangladesh, there is a social norm for shared device access~\cite{sambasivan2018privacy}. This social expectation necessitates features for private browsing or history deletion to maintain privacy within communal usage patterns.
\smallskip

\noindent\textbf{Research Gaps.} Taken together these previous studies on WEIRD populations, they surface a widespread issue of representation across various research domains. The insights applicable to WEIRD populations often do not translate to individuals from different cultural, educational, or economic contexts. Consequently, there is a movement towards adopting more inclusive research methodologies such as incorporating checklists to highlight potential negative impacts~\cite{olteanu2023responsible}. Our study extends these efforts by examining the degree of WEIRD-ness in ICWSM research, contrasting it with findings from two major conferences centered on human-centered design and the democratization and responsibility of AI.

%% file: sections/3_Methodology.tex
\section{Methodology}
\label{sec:methodology}
As researchers, we recognize the importance of offering a \emph{positionality statement} to articulate our viewpoints and situational context within this study. This study is situated in the United Kingdom in the 21\textsuperscript{st} century, writing as authors who primarily work in academia and industry research. We identify as males from Southeast Asia and Southern Europe, and our shared backgrounds include HCI, software engineering, AI, social computing, and urbanism.

In this study, we set out to explore the extent to which social computing papers study WEIRD populations, and in so doing, we formulated three Research Questions (RQs): \\

\noindent \textbf{RQ\textsubscript{1}}: How WEIRD are the datasets in social computing papers (including full research papers, dataset papers, and posters)? \\
\noindent \textbf{RQ\textsubscript{2}:} How WEIRD are the datasets in social computing poster and dataset papers? \\
\noindent \textbf{RQ\textsubscript{3}:} Do cross-country authorships come with
their datasets being less WEIRD? \\

To address these research questions, we focused on ICWSM over other conferences because it is a dedicated community of web and social media researchers, and its specialization in the social computing field guarantees that the research papers under study are comparable. We adopted the methodologies used by \citet{linxen2021weird} and \citet{septiandri2023weird} for measuring the WEIRD-ness in ACM CHI and FAccT conferences. We collected and analyzed 494 papers from the ICWSM conference between 2018 and 2022. This collection of papers included 348 full research papers (70.4\%), 66 dataset papers (13.4\%), and 80 poster papers (16.2\%). Next, we describe a crowdsourcing study designed to collect necessary data for calculating the WEIRD scores of these papers, explain the dataset, define the WEIRD metrics, and discuss any assumptions made in our analysis.

\subsection{Crowdsourcing Study with Full Manual Validation}
\label{subsec:crowdsourcing}
To collect each paper's data for the WEIRD scores' calculation, we conducted a crowdsourcing study on Prolific\footnote{Prolific (\url{https://www.prolific.com/}) is a platform for running crowdsourcing studies, which is known for its high-quality participant pool~\cite{douglas2023prolific}.} with 188 participants, followed by a full manual validation.

\subsubsection{Eligibility Criteria.} 
We defined five eligibility criteria: Human Intelligence Task (HIT) approval rate higher than 95\%; completion of at least 50 HITs; residency in the UK, USA, Ireland, Australia, Canada, or New Zealand (English-speaking countries); fluency in English; and background in Computer Science.

\begin{figure}[t!]
    \centering
    \includegraphics[width=\columnwidth]{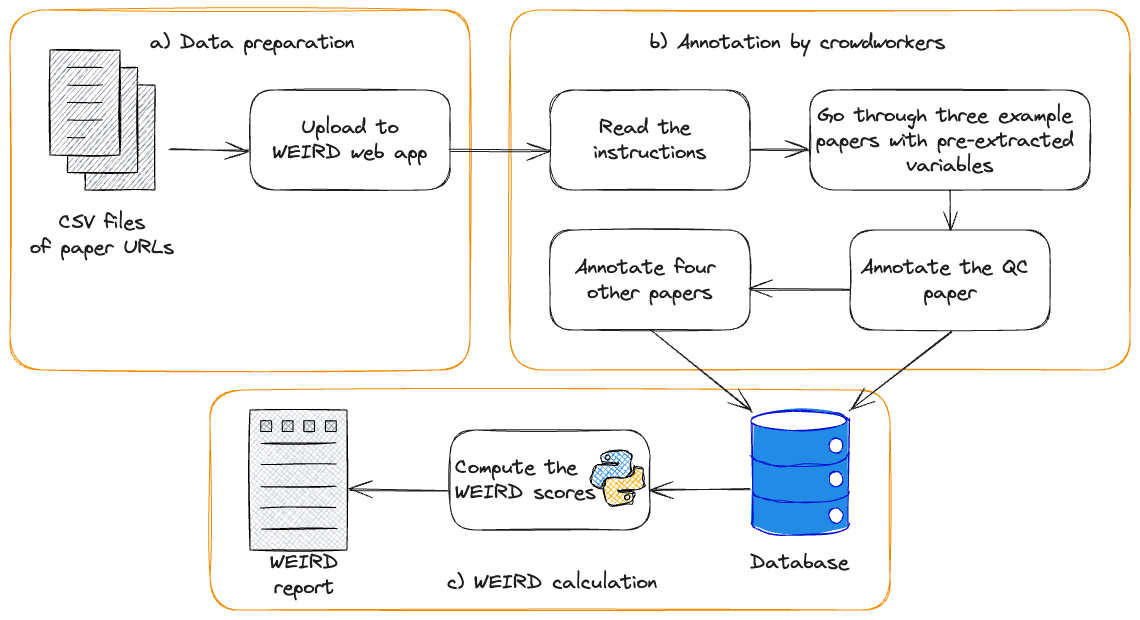}
    \caption{\textbf{Crowdsourcing setup.} \emph{a)} We developed a web app that showed the set of ICWSM papers; \emph{b)} Each crowdworker, having been granted access to the web app through Prolific, annotated five papers. Three example papers were provided to familiarize with the task at hand; and \emph{c)} Based on the obtained data, we computed the WEIRD scores.}
    \label{fig:crowdsourcing}
\end{figure}

\subsubsection{Task Design.} Each participant was presented with five papers on a custom-designed HTML page, which included four research papers and one Quality Control (QC) paper to evaluate the accuracy and reliability of participants' annotations (i.e., extracting data from the provided papers). To ensure high-quality responses, we included two basic attention checks. Participants were instructed to choose `Casablanca' as their favorite city and `Colombia' as their favorite country in order to pass these checks.

The task involved two steps. First, participants familiarized themselves with the task by showing three example papers where variables had already been extracted. These samples acted as a guide, demonstrating the desired approach and style for identifying and annotating variables. Second, participants were tasked with pinpointing and extracting specific variables from the papers such as author affiliations and their countries, the source of the dataset (e.g., a social media platform), and the representation of countries in the dataset (e.g., the count of Twitter users). To guarantee the accuracy of the data collected, variables for each paper were extracted twice.

\subsubsection{Time and Payment.} The estimated time allocated for task completion was around 23 minutes (3 minutes for the initial setup and guidance, and 20 minutes dedicated to the primary task). Consistent with Prolific's minimum wage policy, participants received compensation at a rate of \pounds9 per hour.

\subsubsection{Data Quality.} All 188 participants successfully passed the two attention checks. However, only 38.2\% managed to accurately extract the necessary information for computing the WEIRD scores. It is important to note that we adopted a rather conservative approach, requiring all required fields to be accurately filled for a response to be considered ``high-quality''. Consequently, to ensure the precision of our data, two authors manually reviewed and corrected the annotations for 216 papers.

\subsection{Defining WEIRD Scores}
\label{subsec:weird-score}
This study aims to measure the extent to which ICWSM papers study WEIRD populations. To achieve that, we adopted and extended the methodology and metrics used to explore the WEIRD-ness of CHI and FAccT conferences~\cite{linxen2021weird, septiandri2023weird}. Next, we explain each of the WEIRD scores (Table~\ref{tab:formulae}).

For defining the Western variable, we used Huntington's \textit{Clash of Civilizations} framework~\cite{huntington2000clash}, which posits that global conflicts are primarily driven by cultural differences. The identification of Western civilization is based on cultural and historical factors, including the prevalent language, religious practices, and origins in ancient Greek and Roman traditions—characterized by Christian dominance, the Latin script, and democratic political systems. However, Huntington acknowledges ambiguities in categorizing some nations as ``torn countries''. For example, Turkey, despite its Western-oriented reforms, NATO membership, and EU aspirations, is still classified as non-Western due to its Islamic roots. All European Union members are considered Western~\cite{eu2022country}, whereas countries like Japan, South Korea, Chile, and Argentina, despite meeting criteria of being Educated, Industrialized, Rich, and Democratic, do not fall under the Western category.

For defining the Educated variable, we used the average years of schooling per person (as reported in the UNDP Human Development Report~\cite{undp2022hdr}) to represent a country's education level. This metric measures the average schooling years that adults aged 25 and above have completed. While the OECD's PISA index serves as an alternative, we chose the UNDP's measure due to its reliability in yielding consistent and replicable results, which is in line with findings from previous studies~\cite{linxen2021weird, septiandri2023weird}.

For defining the Industrialized variable, we used the Competitive Industrial Performance (CIP) Index by the United Nations Industrial Development Organization (UNIDO)~\cite{cipi2020unido}. This index evaluates a nation's capability in competitively manufacturing goods. While an alternative indicator, the gross domestic product (GDP) per capita adjusted for purchasing power parity (PPP)~\cite{worldbank2022gdp}, has been used in a previous study~\cite{linxen2021weird}, the CIP Index was selected for its specific focus on industrial performance. GDP per capita is a measure of a country's economic output divided by its population, reflecting the overall wealth produced over a year. However, we noted that GDP per capita closely aligns with gross national income (GNI) per capita, which has been used to measure the `Rich' variable in previous studies.

\begin{table}[t!]
    \centering
     \caption{Toy example of two hypothetical papers that analyze datasets with or without mentioning the country of the users. In each row, the values represent the normalized country representation. If the dataset is taken from a web platform, the country's representation is equal to the platform's penetration rates.}
    \label{tab:example}
    \scalebox{0.8}{
        \begin{tabular}{cll|rrrr}
        \hline
        Dataset & Paper & Description & US & UK & Japan & Mexico \\
        \hline
        1 & A & Computer & 0.80 & 0.00 & 0.20 & 0.00 \\
        2 & A & Twitter & 0.80 & 0.10 & 0.08 & 0.02 \\
        3 & B & Crowdsourced & 1.00 & 0.00 & 0.00 & 0.00 \\
        4 & B & Twitter & 0.00 & 1.00 & 0.00 & 0.00 \\
        \hline
        \hline
        & & $s_c$ & 1.30 & 0.55 & 0.14 & 0.01 \\
        \hline
        \end{tabular}
    }
\end{table}

For defining the Rich variable, we used the gross national income (GNI) per capita, adjusted for purchasing power parity (PPP), as suggested by previous studies~\cite{arnett2008neglected, linxen2021weird}. This measure reflects the average income and standard of living within a country by aggregating the total income produced by its residents and businesses. The GNI per capita is reported in international dollars to allow for equitable comparisons across different nations.

For defining the Democratic variable, we used the ``political rights'' scores provided by Freedom House~\cite{freedom2022countries}, an American non-profit organization focused on research related to democracy, freedom, and human rights. This score reflects the level of political freedoms and rights available to the citizens of a country. While the Democracy Index from the Economist Intelligence Unit (EIU) could serve as an alternative~\cite{democracy2022economist}, we opted for Freedom House's political rights scores to enhance the reproducibility of our findings aligned with previous studies~\cite{linxen2021weird, septiandri2023weird}.

\begin{table*}[t!]
    \centering
    \caption{Formulae to compute the WEIRD variables, adopted from \citet{linxen2021weird} and \citet{septiandri2023weird} and extended to be applicable for social media data. $\mathbb{E}[.]$ indicates the expected value of a random variable, and $\vec{X}$ indicates a vector of value for variable $X$ from all sampled countries.}
    \label{tab:formulae}
    \scalebox{0.75}{
    \begin{tabular}{lllp{12cm}}
        \toprule
        Symbol & Variable & Formula & Description \\
        \midrule
        $c$ & Country & - & Country where the samples are from \\
        $\pi$ & Population & - & Population size of a country based on World Population Prospects 2022 \cite{population2022un} \\
        \midrule
        $W_c$ & Western & $1~\mathrm{if}~c \in \mathrm{Western~else}~0$ & Whether country $c$ is Western based on Huntington classification~\cite{huntington2000clash} \\
        $E_c$ & Educated       & $\mathbb{E}_c[\mathrm{years~of~schooling}]$ & Mean years of schooling for country $c$ based on UNDP Human Development Report (2022)~\cite{undp2022hdr} \\
        $I_c$ & Industrialized & $\mathrm{CIP}_c$ & Level of industrialization for country $c$ based on the Competitive Industrial Performance (CIP) Index from the United Nations Industrial Development Organization (UNIDO)~\cite{cipi2020unido} \\
        $R_c$ & Rich           & $\mathrm{GNI~per~capita}_c$ & Wealth of country $c$ based on World Bank GNI per capita, PPP (current Int\$, 2020)~\cite{worldbank2022gni} \\
        $D_c$ & Democratic     & $\mathrm{political~rights}_c$ & Level of democracy for country $c$ based on Freedom House Political Rights (2022)~\cite{freedom2022countries} \\
        \midrule
        $s_c$ & Total representation per country & - & Sum of fractional representations from all papers studying country $c$ \\
        $\psi_{c}$ & Papers ratio per country & $\frac{s_c / \sum_c s_c}{\pi_c / \sum_c \pi_c}$ & Ratio of the proportion of papers for country $c$ to the proportion of population size for country $c$ \\
        $\tau(., .)$ & Kendall rank correlation & $\frac{P - Q}{\sqrt{(P + Q + T) \cdot (P + Q + U)}}$ & The similarity of two rankings, e.g. $\vec{\psi}$ and $\vec{E}$; $\vec{\psi}$ and $\vec{R}$. $P$ is the number of concordant pairs, $Q$ is the number of discordant pairs, $T$ is the number of ties in the first variable, and $U$ is the number of ties in the second variable. Concordant pairs are pairs of observations in which the two variables are ranked in the same order, while discordant pairs are pairs of observations in which the two variables are ranked in opposite orders~\cite{agresti2010analysis}. \\
        \midrule
        $W$-score & Western score & $\frac{1}{N} \sum_c W_c$ & Expected value of how Western a conference is from all sampled countries \\
        $E$-score & Educated score & $\tau(\vec{\psi}, \vec{E})$ & How correlated papers ratio and mean years of schooling from all sampled countries \\
        $I$-score & Industrialized score & $\tau(\vec{\psi}, \vec{I})$ & How correlated papers ratio and level of industrialization from all sampled countries \\
        $R$-score & Rich score & $\tau(\vec{\psi}, \vec{R})$ & How correlated papers ratio and level of wealth from all sampled countries \\
        $D$-score & Democratic score & $\tau(\vec{\psi}, \vec{D})$ & How correlated papers ratio and level of democracy from all sampled countries \\
        \bottomrule
    \end{tabular}
    }
    
\end{table*}

\subsection{Computing WEIRD on Social Media Data}
\label{subsec:analysis}

\subsubsection{Terminology.} We adopted and extended the metrics reported in \cite{linxen2021weird} and \cite{septiandri2023weird} (Table~\ref{tab:formulae}). For determining the WEIRD scores, we used the Kendall rank correlation~\cite{agresti2010analysis}, which varies between -1 and 1 and measures the strength and direction of association between two variables. A coefficient of 1 signifies a perfect positive correlation, -1 a perfect negative correlation, and 0 no correlation at all. To contextualize these correlations, a coefficient of 1 suggests that ICWSM papers predominantly focus on WEIRD populations, whereas a coefficient of 0 indicates a more balanced representation of study participants relative to global populations. 

\subsubsection{Assumptions.} ICWSM papers often analyze data from social media platforms. However, some papers may not provide all the necessary information about the datasets they are analyzing (typically social media datasets) required to compute the WEIRD score, or they may be analyzing multiple datasets in the same paper. To this end, we made three assumptions to compute the WEIRD scores: (a) each dataset analyzed in a paper is given equal importance; (b) all papers have the same significance within the conference proceedings; and (c) for any dataset that does not explicitly mention the country of the users, we use the platform's penetration rates across countries as the country representation in the dataset (available in our data repository\footnote{\url{https://anonymous.4open.science/r/weird-icwsm-3638}}). However, the use of the platform's penetration rate has its own limitations because it may yield a mapping to countries that are potentially less WEIRD. A more conservative approach would be to run the analysis without the penetration rates, which, as we shall see in the Results section, yields more Western yet less EIRD scores.

To illustrate these assumptions, let us examine two hypothetical papers (Table~\ref{tab:example}). Paper A studies two datasets: the first is a computer network dataset with 800 nodes in the US and 200 nodes in Japan (dataset 1 in Table~\ref{tab:example}), and the second is a Twitter network of 1,000,000 users whose countries are unknown (dataset 2). Paper B studies two other datasets: a dataset from a crowdsourced study in the US with 350 crowdworkers (dataset 3), and another dataset of 10,000 Twitter users in the UK (dataset 4).

To obtain the total representation of country $c$ across all papers in our corpus, we computed the \emph{dataset-level} country representations first, then aggregated those representations to obtain \emph{paper-level} country representations, and, in turn, aggregated those values to obtain the total representation of country $c$. For these computations, we used the data available in our repository; Table~\ref{tab:example} serve as a toy example.

To compute the dataset-level country representation, we calculated the fraction of users in a dataset from country $c$. For example, the Computer Network dataset (dataset 1) contains 800 users from the US and 200 users from Japan (Table~\ref{tab:example}). Therefore, the representation of the US would be 0.8 and that of Japan 0.2, ensuring that the total sum of all countries' representations equals one.

To compute the paper-level country representation, we averaged the representation values of country $c$ for all datasets in the paper. For example, Paper A has a representation value $c$ of 0.8 for the US, and Paper B has a value of 0.5.

Finally, to obtain the overall representation of country $c$ across all papers (which we call $s_c$ in Table~\ref{tab:example}), we summed the paper-level representations across all papers. For example, the $s_c$ value would be 1.3 for the US, 0.55 for the UK, 0.14 for Japan, and 0.01 for Mexico. The higher the $s_c$, the more frequently the country $c$ is studied across papers.

%% file: sections/4_Results.tex
\section{Results}
\label{sec:results}
Out of the 494 papers we collected, we removed those focusing solely on synthetic datasets or those where the country of origin was unclear. This resulted in 420 papers (85\% of the original total) being eligible for analysis.

\begin{figure*}[t!]
  \centering
  \includegraphics[width=.67\textwidth]{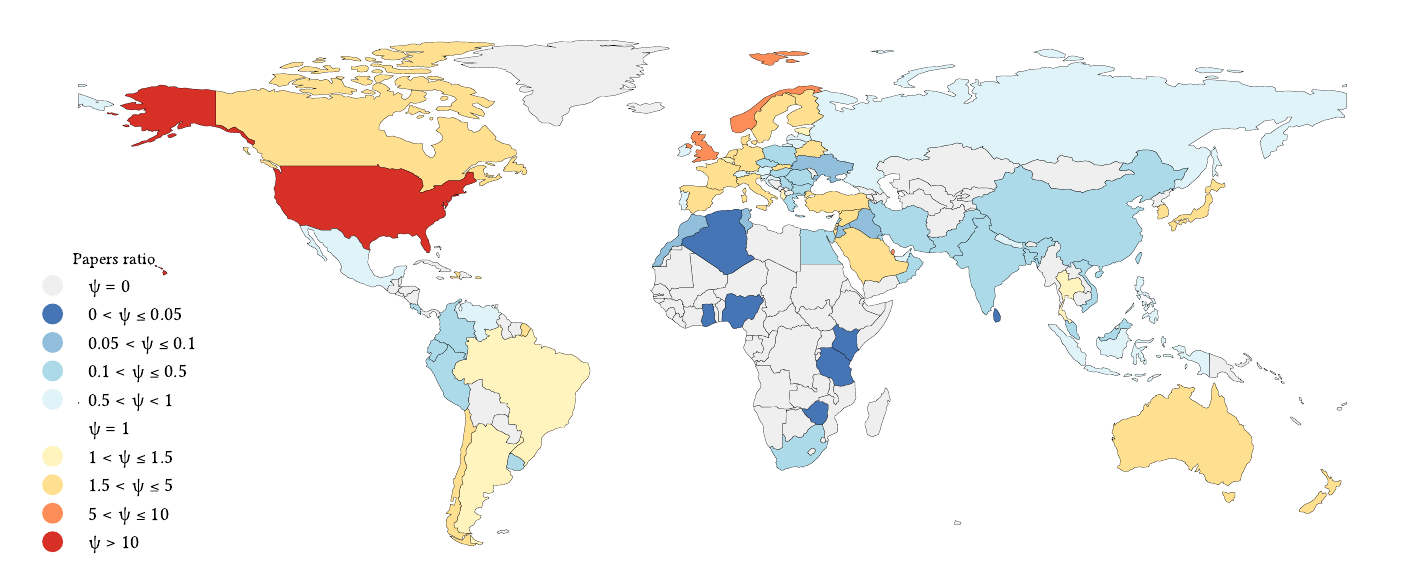}
  \caption{Paper distribution ratio, $\psi_{c}$, reflects the extent of over-representation ($\psi_{c} > 1$) and under-representation ($\psi_{c} < 1$) of countries in ICWSM papers from 2018 to 2022. This ratio is calculated by summing the fractional contributions of each country $c$ across all papers and normalizing it by the population of $c$. Countries not included in the ICWSM papers during the period under study are shown in light gray ($\psi_{c} = 0$), whereas darker shades of blue and red depict countries that are under-represented and over-represented, respectively.}
  \label{fig:map-icwsm}
\end{figure*}

Table~\ref{table:top_countries} presents the top 10 countries based on the paper ratio ($\psi_{c}$) and their overall representation in all papers ($s_c$). The analysis suggests a significant over-representation of the US in the ICWSM community, with an $s_c$ score of 182.66, attributed largely to the focus on social media datasets from English-speaking regions (e.g., Twitter, Reddit, and Facebook). It also highlights that authors from India and Japan frequently conduct studies on communities within their respective countries.

\begin{table*}[t!]
    \centering
    \caption{Between 2018 and 2022, the top 10 countries analyzed in ICWSM papers are ranked based on: $s_c$, the total representation of country $c$; $\%{c}$, the percentage of papers involving country $c$; and $\psi{c}$, the papers ratio representing the overall presence of country $c$ in research, normalized by the country's population size. USA has the highest total representation among all the countries, and this is partly attributed to our method of assigning country data based on media platform penetration rates when specific countries are not mentioned. Since $\psi_{c}$ adjusts for population size, countries with smaller populations (e.g., Dominica, Malta, and French Guiana) tend to appear over-represented.}
    \label{table:top_countries}
    \scalebox{0.9}{
    \begin{tabular}{lrrr|lrrr}
        \toprule
        \multicolumn{4}{c}{Top countries by $\psi_{c}$} & \multicolumn{4}{c}{Top countries by $s_{c}$} \\
        \midrule
        Country & $s_c$ & $\%_{c}$ & $\psi_{c}$ & Country & $s_c$ & $\%_{c}$ & $\psi_{c}$ \\
        \midrule
        Dominica & 0.20 & 0.05\% & 53.96 & United States of America & 182.66 & 45.25\% & 10.56 \\
        Malta & 1.00 & 0.25\% & 37.69 & India & 25.92 & 6.42\% & 0.36 \\
        United States of America & 182.66 & 45.25\% & 10.56 & Japan & 21.77 & 5.39\% & 3.38 \\
        French Guiana & 0.12 & 0.03\% & 8.34 & United Kingdom & 18.65 & 4.62\% & 5.40 \\
        Singapore & 2.31 & 0.57\% & 7.59 & Brazil & 13.64 & 3.38\% & 1.24 \\
        Qatar & 1.03 & 0.26\% & 7.27 & China & 12.69 & 3.14\% & 0.17 \\
        Barbados & 0.09 & 0.02\% & 6.37 & Germany & 9.82 & 2.43\% & 2.29 \\
        United Kingdom & 18.65 & 4.62\% & 5.40 & Indonesia & 9.19 & 2.28\% & 0.66 \\
        Norway & 1.44 & 0.36\% & 5.21 & France & 8.14 & 2.02\% & 2.45 \\
        Israel & 2.15 & 0.53\% & 4.77 & Italy & 7.18 & 1.78\% & 2.35 \\
        \bottomrule
    \end{tabular}
    }
\end{table*}

\subsection{RQ1: How WEIRD are the datasets in social computing papers (including full ICWSM research papers, dataset papers, and posters)?}
\label{sec:result-rq1}

We found that 37\% of ICWSM papers focus solely on Western countries (Table~\ref{table:western}). Conversely, a considerable portion of the papers (51\%) examine a combination of Western and non-Western countries, likely reflecting our methodological assumptions. This mix of countries in the studies is significantly higher compared to FAccT and CHI, potentially up to six times greater. Even with a larger share of studies featuring a diverse set of countries, 27\% of the ICWSM papers are dedicated to data exclusively from the US, a percentage that remains lower than that of FAccT (93.6\%) and CHI (54.8\%).

\begin{table*}
    \centering
    \caption{The representation of countries in ICWSM papers varies, with FAccT and CHI papers primarily involving participants from Western countries (numbers were taken from \citet{septiandri2023weird} and \citet{linxen2021weird}). Additionally, ICWSM showcases a broader array of nationalities, which is likely driven by the use of social media data.}
    \label{table:western}
    \scalebox{0.8}{
    \begin{tabular}{lrr|rr|rr}
        \toprule
        {} & \multicolumn{2}{c}{\textbf{ICWSM}} & \multicolumn{2}{c}{\textbf{FAccT} (\citeauthor{septiandri2023weird})} & \multicolumn{2}{c}{\textbf{CHI} (\citeauthor{linxen2021weird})} \\
        Variable      &  n   & \%      &  n    &      \% &  n    &      \% \\
        \midrule
        Exclusively Western     & 157 & 37.38\% & 108 & 84.38\% &  817 & 75.93\% \\
        Exclusively non-Western &  51 & 12.14\% &   9 &  7.03\% &  197 & 18.31\% \\
        Mixed                   & 212 & 50.48\% &  11 &  8.59\% &   62 &  5.76\% \\
        \midrule
        Total                   & 420 &  100\%  & 128 & 100\%   & 1076 & 100\%   \\
        \bottomrule
    \end{tabular}
    }
\end{table*}

ICWSM positions itself between FAccT and CHI in terms of the Educated and Rich variables (Table~\ref{table:weird_index}). Yet, it scores lowest on the Democratic variable and highest on the Industrialized variable when compared to the other two conferences. Although FAccT's Democratic score surpasses ICWSM's by a mere 0.05, ICWSM's Industrialized score of 0.35 is considerably higher than FAccT's.\footnote{The figures for FAccT and CHI were recalculated using the datasets made available by \citet{septiandri2023weird} and \citet{linxen2021weird} under the MIT license.} Additionally, we calculated the 95\% confidence intervals (CI) for each correlation coefficient using the bootstrapping method.

Among the 420 papers we analyzed, 181 did not explicitly specify the users' country. After removing these 181 papers (for which we used penetration rates) from our analysis, we found that the 239 remaining papers primarily examined Western samples and recorded lower EIRD scores. This suggests that our main conclusions remained the same, even when adjusting for penetration rates in our analysis.

\begin{table*}[t]
    \centering
    \caption{Kendall rank correlations ($\tau$) between the papers ratio $\psi_{c}$ and the Educated, Industrialized, Rich, Democratic scores, with confidence intervals derived from 10,000 bootstrap samples. Levels of significance: *$p < .05$, **$p < .01$, ***$p < .001$.}
    \label{table:weird_index}
    \scalebox{0.8}{
    \begin{tabular}{lll|ll|ll}
        \toprule
        {} & \multicolumn{2}{c}{\textbf{ICWSM}} & \multicolumn{2}{c}{\textbf{FAccT}} (\citeauthor{septiandri2023weird}) & \multicolumn{2}{c}{\textbf{CHI} (\citeauthor{linxen2021weird})} \\
        Variable       &  $\tau$ &   95\% CI $\tau$ &  $\tau$ &   95\% CI $\tau$ &  $\tau$ &   95\% CI $\tau$ \\
        \midrule
        Educated       & 0.36*** & [0.23, 0.49] &  0.31**  &   [0.12, 0.50] & 0.43*** &   [0.30, 0.57] \\
        Industrialized & 0.35*** & [0.20, 0.50] &  0.01    &   [-0.21, 0.23] & 0.27*** &   [0.13, 0.41] \\
        Rich           & 0.49*** & [0.36, 0.61] &  0.34*** &   [0.20, 0.49] & 0.50*** &   [0.37, 0.62] \\
        Democratic     & 0.32*** & [0.20, 0.45] &  0.37*** &   [0.20, 0.55] & 0.51*** &   [0.41, 0.61] \\
        \bottomrule
    \end{tabular}
    }
\end{table*}

\subsection{RQ2: How WEIRD are the datasets in social computing poster and dataset papers?}
\label{sec:rq2}

\begin{table*}[t]
    \centering
    \caption{Kendall rank correlations ($\tau$)  between the papers ratio $\psi_{c}$ and the Educated, Industrialized, Rich, Democratic (EIRD) scores, comparing full papers to dataset and poster papers. The differences in EIRD scores ($\Delta\tau$) are calculated alongside confidence intervals obtained from 10,000 bootstrap samples. Levels of significance: *$p < .05$, **$p < .01$, ***$p < .001$.}
    \label{table:full-vs-rest}
    \scalebox{0.85}{
    \begin{tabular}{lll|ll|ll}
        \toprule
        {} & \multicolumn{2}{c}{\textbf{Difference}} & \multicolumn{2}{c}{\textbf{Full Papers}} & \multicolumn{2}{c}{\textbf{Dataset \& Poster Papers}} \\
        Variable       &  $\Delta\tau$ &   95\% CI $\Delta\tau$ &  $\tau$ &   95\% CI $\tau$ &  $\tau$ &   95\% CI $\tau$ \\
        \midrule
        Educated       & 0.12* & [-0.13, 0.11] & 0.40*** & [0.27, 0.52] & 0.28*** & [0.14, 0.41] \\
        Industrialized & 0.08* & [-0.15, 0.10] & 0.39*** & [0.24, 0.53] & 0.30*** & [0.15, 0.46] \\
        Rich           & 0.07 & [-0.14, 0.15] & 0.48*** & [0.36, 0.61] & 0.41*** & [0.28, 0.54] \\
        Democratic     & 0.18** & [-0.14, 0.12] & 0.39*** & [0.27, 0.51] & 0.21*** & [0.06, 0.36] \\
        \bottomrule
    \end{tabular}
    }
\end{table*}

One may also hypothesize that other publication tracks at ICWSM such as dataset and poster papers might contribute to a broader diversity in the samples studied. The null hypothesis suggests no difference in EIRD scores between full research papers and those submitted as dataset or poster papers. To test this hypothesis, we used a permutation test based on the premise that all permutations are equally probable. We found significant disparities in the Educated and Democratic variables (Table~\ref{table:full-vs-rest}), suggesting that dataset and poster papers often involve samples from less Educated and less Democratic nations. This pattern may reflect the ICWSM community's inclination to engage with political subjects across a spectrum of democracies, including discussions on events in both democratic countries such as Brexit~\cite{calisir2020long} and the US election~\cite{abilov2021voterfraud2020} and less democratic contexts such as the presidential elections in Belarus~\cite{hohn2022belelect}, the political crisis in Brazil~\cite{oliveira2018politicians}, and censorship challenges in Turkey, India, and Russia~\cite{elmas2021dataset}. While studies concerning US politics commonly appear in the main track (i.e., full research papers), the aforementioned examples were predominantly found in dataset or poster papers.

To explore the relationship between the Western score and the type of paper (full research papers vs. dataset and poster papers), we conducted a chi-squared contingency test. We found no statistically significant difference in the Western-centric focus of the studies across these paper types ($\chi^2(2, N = 420) = 3.84, p > .05$).

\subsection{RQ3: Do cross-country authorships come with their datasets being less WEIRD?}
\label{sec:rq3}

\begin{table*}[tbp]
    \centering
    \caption{For ICWSM, the Pearson correlation coefficient ($\rho$) indicates a negative correlation between the number of unique author affiliation countries and the Educated and Democratic scores. In contrast, for FAccT, there is no association with any of the EIRD scores, whereas CHI shows a negative correlation across all EIRD scores.}
    \label{tab:affiliations}
    \scalebox{0.85}{
    \begin{tabular}{lcc|cc|cc}
        \toprule
        {} & \multicolumn{2}{c}{\textbf{ICWSM}} & \multicolumn{2}{c}{\textbf{FAccT}} (\citeauthor{septiandri2023weird}) & \multicolumn{2}{c}{\textbf{CHI} (\citeauthor{linxen2021weird})} \\
        Variable & $\rho$ & 95\% CI & $\rho$ & 95\% CI & $\rho$ & 95\% CI \\
        \midrule
        Educated & -0.11 & [-0.20, -0.01] & -0.06 & [-0.23, 0.11] & -0.22 & [-0.28, -0.16] \\
        Industrialized & -0.03 & [-0.12, 0.07] & -0.05 & [-0.22, 0.12] & -0.21 & [-0.27, -0.15] \\
        Rich & -0.06 & [-0.16, 0.04] & -0.06 & [-0.23, 0.11] & -0.23 & [-0.28, -0.17] \\
        Democratic & -0.16 & [-0.26, -0.07] & 0.05 & [-0.13, 0.22] & -0.11 & [-0.17, -0.05] \\
        \bottomrule
    \end{tabular}
    }
\end{table*}

We also examined how the geographic diversity of authors influences the diversity of samples in research papers. In particular, we investigated the premise that papers written by authors from different countries are less likely to focus on WEIRD samples, given that the authors' institutional affiliations frequently mirror the demographics of their research subjects. Typically, researchers affiliated with US institutions focus on US participants, while those affiliated with Asian institutions tend to include Asian participants in their studies. This pattern may contribute to uniformity in sample selection, potentially skewing research findings.

To quantify this relationship, we identified the countries of authors' affiliations using metadata from  OpenAlex API.\footnote{https://docs.openalex.org/} For authors associated with more than one institution, only their primary affiliation was considered. All 420 papers that we analyzed featured at least one author with a country of affiliation, with a median author count of $4.0$ ($\sigma = 2.0$). We found that authors come from 43 distinct countries, 54\% of which are considered Western, and that 72\% of the papers were authored by teams from just one country.

Next, we explored the relationship between the diversity of author affiliations by country and the WEIRD scores using Pearson correlation coefficients. We found a statistically significant negative correlation between the diversity of countries authors are affiliated with and both the Educated and Democratic scores (Table~\ref{tab:affiliations}). This suggests that papers authored by individuals from multiple countries tend to focus on less Educated or Democratic countries. For example, studies on China featuring collaborations of at least 6 authors (e.g., those exploring urban-rural stress differences~\cite{cui2022social}, anti-China sentiment~\cite{shen2022xing}, and the dynamics of the social commerce site Beidian~\cite{cao2020your}), often include authors not primarily based in China. This trend deviates from what is observed at CHI, where significant correlations were noted across all EIRD metrics, indicating that CHI's cross-national collaborations are linked with samples from countries that score lower across all four EIRD dimensions, not just in terms of Education and Democracy.

Finally, we looked into how authorship across different countries (classified as either exclusively Western, exclusively non-Western, or a combination) correlates with the geographical focus of their studies. To test this, we used a chi-squared contingency test. The test yielded a chi-squared statistic of 23.86 with 4 degrees of freedom ($p < 0.001$). Such a finding strongly suggests that the association between the authors' geographical origins and the focus of their research is not coincidental but indicates a genuine link. This supports the conclusion drawn by \citet{linxen2021weird}'s study, emphasizing the importance of fostering collaborations between Western and non-Western countries to mitigate the WEIRD bias in CHI research.

\subsection{Papers low in Western and Democratic Variables}
\label{sec:low_w_d}
Overall, it was evident that ICWSM papers have significantly lower values in the Western (W) variable compared to those of CHI and FAccT. To illustrate this empirical finding, we manually inspected three papers with the lowest W scores. For example, the paper \emph{``BelElect: A New Dataset for Bias Research from a 'Dark' Platform''} presents a new Telegram corpus in Russian and Belarusian languages tailored for research on linguistic bias in political news; the paper \emph{``Know It to Defeat It: Exploring Health Rumor Characteristics and Debunking Efforts on Chinese Social Media during the COVID-19 Crisis''} conducted an analysis of data obtained from Weibo, a Chinese microblogging site; the paper \emph{``DISMISS: Database of Indian Social Media Influencers on Twitter''} presented a systematically categorized database of influential accounts on Twitter in India. Similarly, we found that poster and dataset papers tended to focus on samples from countries with lower levels of democracy (D). We manually inspected three papers with the lowest D scores. For example, the paper \emph{``BelElect: A New Dataset for Bias Research from a ”Dark” Platform''} proposed a new corpus in Russian and Belarussian languages tailored for research on linguistic bias in political news; the paper \emph{``MMCHIVED: Multimodal Chile and Venezuela Protest Event Data''} used the the MMCHIVED dataset to study protest events in Chile and Venezuela; and the paper \emph{``FactDrill: A Data Repository of Fact-Checked Social Media Content to StudyFake News Incidents in India''} studied fake news in India.

%% file: sections/5_Discussion.tex
\section{Discussion}
\label{sec:discussion}
By analyzing 420 ICWSM papers published between 2018 and 2022, we found that 37\% of them, irrespective of their type (i.e., full research paper, dataset or poster paper), exclusively study datasets from Western countries. This proportion is  lower than the figures reported in CHI (76\%) and FAccT (84\%) papers. The disparity could be attributed to the diverse range of datasets available both from Western and non-Western countries. Nonetheless, ICWSM papers predominantly focus on samples from countries characterized by higher levels of education, industrialization, and affluence compared to those published in FAccT. This trend may be influenced by the prevalence of datasets from platforms such as Twitter and Reddit, where analyses typically target English-speaking nations. Upon closer examination, we found that dataset and poster papers tend to exhibit lower ``EIRD'' scores (i.e., Educated, Industrialized, Rich, and Democratic) compared to full research papers. Finally, our findings indicated that collaborations spanning multiple countries are correlated with research studies analyzing samples from countries with lower levels of democracy.

\subsection{Implications}
From a theoretical standpoint, our study contributes to the growing body of literature on WEIRD, situated within the broader landscape of Critical Computing literature~\cite{comber2020announcing}. While ICWSM papers demonstrate a more diverse array of dataset origins compared to CHI and FAccT, we echo concerns that research remains predominantly Western- and US-centric~\cite{laufer2022four}. Our study delves into this discourse by not only shedding light on the WEIRD characteristics of ICWSM and juxtaposing them with two other conferences but also by examining differences across conference tracks and exploring the interplay between cross-country authorship and WEIRD metrics. Of particular interest is the ``Democratic'' variable, which measures the degree of political freedoms and rights within a nation. Throughout our analysis, it became evident that studying social media data helps to uncover findings from less democratic countries (e.g., Belarus, China, Syria).

From a practical standpoint, we made three recommendations: \emph{expanding the paper checklist}, \emph{including Responsible AI statements in papers}, and \emph{championing author diversity.}

\subsubsection{Expanding the Paper Checklist.} We propose expanding the current mandatory paper checklist for ICWSM papers, particularly section 5, by incorporating three additional questions. These questions can be phrased in the form of statements, asking authors to: \emph{1)} specify the country of origin of the subjects studied in the dataset; \emph{2)} identify the (social media) platforms from which the data was obtained; and \emph{3)} state the country of origin of the authors who contributed to the paper.
By integrating these questions with the data we have already collected from official sources (and made publicly available) such as GDP, GNI, CIP, and political rights, it becomes possible to develop a scoring system that assesses a paper's ``WEIRD-ness''. However, it is important to note that such reporting is intended to increase awareness and should not be considered a deciding factor in a paper's acceptance.

\subsubsection{Including Responsible AI Statement.} The concept of WEIRD is interconnected with the broader context of developing and deploying fair, transparent, and accountable AI, also known as Responsible AI~\cite{tahaei2023sig}. Incorporating a responsible AI statement is crucial in addressing the potential harms and risks associated with datasets originating from specific countries. Several conferences such as NeurIPS~\cite{ashurst2020guide} and ICML have already begun mandating statements that disclose ``any risks associated with the proposed methods, methodology, application or data collection and data usage''. Recently, \citet{olteanu2023responsible} further emphasized the importance of including impact statements in responsible AI research. These statements aim to reveal any potential negative consequences and contribute to more inclusive research practices. We support these initiatives and encourage the ICWSM community to adopt them as well.

\subsubsection{Championing Author Diversity and Shadow Mentoring Programs.} 
Promoting diversity among authors and fostering collaborations can be achieved through scholarships and shadow mentoring schemes. These initiatives encourage a mix of expertise and gender among researchers~\cite{pfund2006merits}, which can lead to cross-country collaborations and create career development opportunities for underrepresented groups~\cite{talbert2021challenges}. Financial support from non-governmental organizations and foundations could play a significant role in enabling research activities in under-represented nations. Another practical approach is to establish shadow mentoring programs that match senior community members with authors who infrequently publish in the main proceedings. Our findings indicate that authors who predominantly contribute datasets and poster papers often focus on less commonly studied samples, yet their work may not reach the main conference proceedings. Such mentorship could enhance their visibility within the broader publication ecosystem. Moreover, the community could further recognize research on developing nations, possibly through awards or special sessions in the main conference track, instead of limiting these studies to dataset or poster papers.

\subsection{Limitations and Future Work}
Our study comes with six limitations that call for future research efforts. First, the datasets used in ICWSM papers are often based on social media and typically do not specify the rationale behind the choice of countries analyzed. A potential approach to address this issue involves correlating these datasets with the social media usage rates in different countries. Yet, this method might still perpetuate biases, especially when dataset selection is swayed by the dominance of English-language content. Future research should explore the feasibility of using countries with English as a predominant language to more accurately measure social media penetration. Second, the identification of WEIRD populations in our analysis does not suggest a uniformity in perspectives. Even within a single country, many different backgrounds and cultures from immigrants and minority groups add to the variety of views and experiences. Third, to capture the Democratic variable, we relied on the political rights scores from the Freedom House, which receives most of its funding from the US government. This could have introduced a bias in our analysis as every country has its own culture and way of practicing democracy. Future research could explore alternative ways of quantify the Democratic variable such as the Democracy Index from the Economist Intelligence Unit~\cite{democracy2022economist}. While the political scores from the Freedom House focus on observed political rights and civil liberties, the Democracy Index from the Economist Intelligence also considers political participation and culture, providing a wider view of democracy. Fourth, the concept of WEIRD as a measure of inclusivity is incomplete. It overlooks essential aspects such as gender and sexual diversity, racial and ethnic differences, age groups, disabilities, and a range of ideological beliefs~\cite{seaborn2023weird}. Future endeavors should examine alternative concepts such as the WILD (Worldwide, In-situ, Local, and Diverse) to embrace a broader spectrum of inclusivity. Fifth, the date ranges between ICWSM papers and those from previous analyses on CHI and FAccT have some overlap, but they are not identical. We opted in for those date ranges for comparability reasons with previous works. Future studies could collect additional data to align the date ranges. Finally, we studied how WEIRD is social computing research by examining the proceedings of ICWSM. Despite ICWSM being the premier venue that is dedicated to publishing social computing research, future studies could expand our findings by including venues that, in addition to their main themes, also publish social computing research (e.g., the Web Conference or Computer-Supported Cooperative Work and Social Computing). However, two primary issues need to be addressed: 1) establishing selection criteria for relevant papers as we aim to identify a limited subset from the proceedings of those conferences; and 2) developing the capability to analyze all papers from different conferences in a standardized manner.

%% file: sections/6_Conclusion.tex
\section{Conclusion}
\label{sec:conclusion}
By analyzing 420 ICWSM papers published between 2018 and 2022, we observed a surprising trend in the representation of WEIRD samples. Only 37\% of these papers were centered on Western populations, a figure considerably lower than those found in CHI and FAccT. This indicates that ICWSM research tends to include a wider range of datasets from both Western and non-Western regions. Nevertheless, there appears to be a tendency towards studies from more affluent, industrialized countries, likely influenced by the prevalence of datasets from English-speaking regions.